\documentclass[a4paper,12pt]{article} 
\input{epsf}
\newcommand{\bcen}{\begin{center}}
\newcommand{\ecen}{\end{center}}
\newcommand{\btab}{\begin{tabular}}
\newcommand{\etab}{\end{tabular}}
\newcommand{\bdes}{\begin{description}}
\newcommand{\edes}{\end{description}}

\newcommand{\beq}{\begin{equation}}
\newcommand{\eeq}{\end{equation}}
\newcommand{\bea}{\begin{eqnarray}}
\newcommand{\eea}{\end{eqnarray}}
\newcommand{\non}{\nonumber}

\newcommand{\half}{\frac{1}{2}}
\newcommand{\bary}{\begin{array}}
\newcommand{\eary}{\end{array}}
\newcommand{\beps}{\mbox{\boldmath $ \epsilon $}}

\newcommand{\bn} { \mbox{\boldmath $n$}}

\newcommand{\bu} { \mbox{\boldmath $u$}}

\newcommand{\D}[1]{\mbox{d}{#1}} 

\newcommand{\prn}[1] {(\ref{#1})}

\newcommand{\fig}[1]{fig.~\ref{#1}}

\usepackage{harvard}
\usepackage{epsfig}
\usepackage{amssymb}

\newcommand{\Ke}{K_{eff}}

\newcommand{\figcap}[1]{\caption{#1}}

\newcommand{\mytitle}{Pattern Formation in a Substrate--Contactor
System with Two Interacting Incompressible Elastic Films}

\begin{document}
\baselineskip=18pt
%TITLEPAGE AT END

\date{}

\title{\sl {\Large \mytitle}} 
\author{Vijay Shenoy$^1$\footnote{Corresponding author. Tel:
+91-512-597307, Fax:+91-512-597408, email: {\tt vbshenoy@iitk.ac.in}} ~ and Ashutosh Sharma$^2$ \\ Departments of
$^1$Mechanical and $^2$Chemical Engineering\\
Indian Institute of Technology Kanpur\\
UP 208 016, India}
\maketitle

\centerline{{\bf Keywords}: thin films in contact, van der Waals
interaction,   stability and bifurcation
}

\begin{abstract}
The surface stability of two interacting (for example, by van der Waals
forces)  incompressible thin films,
one bonded to a substrate and the other to a contactor, is studied
extending the work of Shenoy and Sharma, {\em Physical Review Letters}
{\bf 18}, 119--122 (2001). The analysis indicates that the wavelength
of the instability depends strongly on the shear moduli and
thicknesses of the films but not on the nature and magnitude of the
interaction. When the films have equal shear moduli, the wavelength of
the instability has an intermediate value between the wavelengths of
the instabilities had each of the films been interacting with rigid
contactors. On the other hand, if the films have different shear
moduli but equal thicknesses, then the wavelength of the instability
is identical to that had the films been interacting with rigid
contactors. In the more general case when the two films have different
shear moduli and thicknesses, the nature of the critical wavelength is
more complex.  When ratio of the shear moduli of the contacting film
to that of the film bonded to the substrate exceeds 5.32, the
wavelength of the of the instability jumps from the value close to
that determined by the thickness of the film bonded to the substrate
to that of the contacting film, as the thickness of the contacting film
is increased.
\end{abstract}

\section{Introduction}
\label{Intro}

The mechanics of adhesion and contact  between two elastic
bodies has attracted attention of researchers over the
years. Following the classical work of Hertz, an important step in
this area was taken by \citeasnoun{Johnson1971} who identified the
importance of the interactions between the contacting surfaces. These
interactions can be due to van der Waals forces, electrostatic forces
etc.~ between the contacting surfaces. Indeed, measurements of van der
Waals forces between surfaces \cite{Tabor1969} were available when
\citeasnoun{Johnson1971} formulated their famous JKR theory of
contact.

More recently, two groups \cite{Ghatak2000,Monch2001} have reported
experiments aimed at understanding contact and adhesion mechanics
between two elastic bodies with planar topology.  The experiments
reported by \citeasnoun{Ghatak2000} used a configuration where the
glass plate was placed in contact with the elastomeric film between
two spacer bars creating a small gap between the film and the glass
plate. \citeasnoun{Monch2001} performed experiments that involved the
contact of a glass plate with an elastomeric thin film bonded rigidly
to a glass substrate.  In both these experimental works it was
observed that the film lost planarity when the contacting glass plate
reached contact proximity (10-50nm), resulting in a pattern with a
well defined wavelength.  The two key features observed in both sets of
experiments are (a) the wavelength of the pattern depends linearly on
the thickness of the film (b) the wavelength is not affected by the
magnitude and nature of the interactions.

The theoretical interpretation of this results were first reported by
\citeasnoun{Shenoy2001} who argued that the instability occurs due to
a competition between the interaction energy of the film with the
contactor and the elastic energy in the film due to inhomogeneous
deformation. They showed that instability sets in when the
ratio of the ``stiffness of interaction'' and the elastic stiffness
(defined as shear modulus divided by the thickness) exceeds a critical
value. In addition, they pointed out that the wavelength of the
instability is determined solely by the elastic energy of the film and
explained the linear dependence of the wavelength on the thickness of
the film. A detailed account of this work can be found in
\citeasnoun{Shenoy2001a}. The origins and nature of this instability
are different from those known to exist in solid films
\cite{Asaro1972,Grinfeld1993,Srolovitz1989,Ramirez1989} and fluid
films \cite{Herminghaus1998,Sharma1998,Reiter2000,Schaffer2000}.

The purpose of this paper is to extend the work of
\citeasnoun{Shenoy2001} to the case when the contactor also has a film
bonded to it as shown in \fig{tfscheme}. The contacting film may have
different properties in that the shear modulus and thickness may be
different from the film bonded to the substrate. It is shown that one of
key features i.~e., the independence of the wavelength of the
instability on the nature of interaction is unchanged. The critical
wavelength, although does not depend on the nature and strength of the
interaction, does depend strongly on the thicknesses and moduli of the
film.  It is shown that the relative stiffnesses of the films play an
important role in the determination of the wavelength of the
instability when the films have widely differing properties;  the
complete 
dependence of the wavelength on the thicknesses and moduli of the film is
obtained.  This paper is written with a purpose of motivating further
experiments in the kinds of systems indicated in \fig{tfscheme} and
is organised as follows. The next section contains a stability analysis
of  two interacting films. Results of the analysis are presented
and discussed in section \ref{Results}. The findings of this paper are
summarised in section \ref{Summary}. The appendix at the end of the paper
contains a short calculation of the normal traction along the surface
of an incompressible film subjected to a sinusoidal deformation.

\section{Stability of Two Interacting Films}
\label{Stability}

\subsection{Model Description}
\label{Model}

A schematic depiction of model considered for the study of stability
of interacting thin films is shown \fig{tfscheme}. The system consists
of a substrate-contactor configuration where both the substrate and
the contactor have thin films (possibly of different materials) bonded
to them. The substrate film (called film $a$) has a thickness $h_a$,
while the contactor film (called film $b$) has a thickness $h_b$. A
coordinate system described by coordinates $(x_1,x_2)$ is used to
describe position vectors. The positive `2' direction is the outward
normal of film $a$, while the negative `2' direction point in the
normal direction to the surface of the film $b$. We consider only
plane strain deformation of the system.

The equilibrium configuration of this system is determined by
the potential energy
\bea
\Pi =   \int_V W(\beps) \D{V}  - \int_S U((\bu^a -\bu^b) \cdot \bn) \, \D{S}  \label{ex_energy}
\eea
where $W$ is the strain energy density, $\beps$ is the strain tensor,
$\bu$ is the displacement vector (the superscript $a$ denotes value in
film $a$ etc.) with $V$ being an appropriate measure of the volume
(includes volumes of both the films), and $S$ is the interfacial area
of the two films over which they interact. The function $U$ represents the interaction
potential between the two films; it is this term that gives rise to
interesting physics in this system. We make two key physical
assumptions in writing the total potential energy: (i) the
contribution from the surface energies of the films is negligible  (ii)
the films are considered to be made of incompressible elastic
materials. Both of these assumptions are valid in physical
systems where such instabilities are triggered, as shown by
\cite{Shenoy2001}. 

The physical origins of such an interaction can be any of the several
causes -- van der Waals forces, electrostatic forces between
films etc; the potential $U$ is a generic interaction potential.  If
the potential is due to the attractive van der Waals interaction, then
$U$ is described by
\bea
U((\bu^a - \bu^b) \cdot \bn) = \frac{1}{12 \pi} \frac{A}{(d_o - (\bu^a
- \bu^b) \cdot \bn)^2} \label{vanderWaals}
\eea
where $A$ is the Hamaker constant (of the order of $10^{-19} J$), and
$d_o$ is the distance between the surfaces of the two films, i.~e.,
the gap thickness. The strength of the interaction potential is
determined by the gap thickness $d_o$. The contactor is imagined to be
brought in the proximity of the substrate by reducing the distance
$d_o$ -- the interesting quantity being the distance $d_c$ at which
the attractive interactions are strong enough to trigger instability
in the system.

\begin{figure}
\centerline{\input{tfscheme.pstex_t}}
\figcap{A thin elastic film bonded to a rigid substrate interacting
with another film bonded on to a  contactor. The dashed lines shows
possible inhomogeneous deformation of the films when instability sets
in. (Distances are not to scale.) }
\label{tfscheme}
\end{figure}

Linear stability analysis is performed using linear kinematics and a
linearised interaction potential. To this end, the interaction
potential is  expanded in a power series about the reference state of
the undeformed films and terms of up to quadratic order in $(\bu^a -
\bu^b) \cdot \bn$ are retained
\bea
U((\bu^a - \bu^b) \cdot \bn) \approx U_0 + F \, (\bu^a -\bu^b) \cdot \bn
+ \half Y \, ( (\bu^a - \bu^b) \cdot \bn)^2 
\label{uapprox} 
\eea
where
\bea
U_0 = U(0), \;\;\;\;\; F = U'(0), \;\;\;\;\; Y = U''(0).
\eea
The quantity $Y$, called the {\em interaction stiffness}, is of importance and governs the stability of the
system. The aim of the analysis is to find the condition(s) on $Y$
under which instability sets in.

The above approximations give an expression for the approximate
potential energy as 
\bea
\Pi_a = \int_V W(\beps) \D{V}  
- \int_S \left( U_0 + F \, (\bu^a - \bu^b) \cdot\bn
+ \frac{1}{2} Y \, ((\bu^a - \bu^b) \cdot \bn)^2 \right) \, \D{S}. \label{ap_energy}
\eea 
The strain energy density $W$ is 
\bea
W(\beps)  = 
\left\{ 
\begin{array}{c}
\displaystyle{\frac{\mu_a}{2}} \, \beps : \beps \;\;\;\;\;\;\;
\mbox{in film $a$} \\ \\
\displaystyle{\frac{\mu_b}{2}} \, \beps : \beps \;\;\;\;\;\;\; \mbox{in film $b$}
\end{array} 
\right.
\eea
where $\mu_a$ and $\mu_b$ are respectively the shear moduli of the films
$a$ and $b$.                                                          
The equilibrium displacement fields in the films minimise the
potential energy \prn{ap_energy} while satisfying the rigid boundary
conditions at the film-substrate interface in film $a$, and the film
contactor interface in film $b$. In addition, the stresses derived
from these displacement fields satisfy the condition of vanishing
shear stress $\sigma_{12}$ at the surface in both the films. The
normal stresses satisfy the condition
\bea
\sigma_{22}^a  & = &   F + Y(u^a_2 - u^b_2) \\
\non \\
\sigma_{22}^b  & = &   F + Y(u^a_2 - u^b_2) 
\eea
at  their respective surfaces.

\noindent
{\em Homogeneous Solution:} Since both the films are incompressible,
the homogeneous solution has displacements vanishing everywhere
 in both the films, and
stress state in both the films is one of constant pressure (equal to
$F$). The point in question is the stability of this homogeneous
solution, i.~e., for what value(s) of $Y$ does instability occur?

\subsection{Stability Analysis}
To study the stability of the system above, the homogeneous solution
is perturbed by bifurcation fields (denoted by $u_i$ here and henceforth)
such that the surfaces of the film $a$ and $b$, respectively, have
displacements 
\bea
u_2^a(x_1) & = & \alpha \cos(k x_1) \label{ua} \\
\non \\
u_2^b(x_1) & = & \beta \cos(k x_1) \label{ub}. 
\eea
The {\em additional stresses} produced by these fields satisfy
\bea
\sigma_{12}^a = 0, \;\;\;\; \sigma_{12}^b = 0, \label{noshear}
\eea
and
\bea
\sigma_{22}^a  =  Y(u^a_2 - u^b_2) \label{sabif} \\
\non \\
\sigma_{22}^b  =  Y(u^a_2 - u^b_2) \label{sbbif} 
\eea
along the interacting surfaces of the films. All symbols have obvious meanings.

It is shown in the appendix that the stress $\sigma^a_{22}$ at the
surface of a film $a$ whose displacement is $u_2^a(x_1) = \alpha \cos(k
x_1)$ with $\sigma^a_{12} =0$ is given by
\bea
\sigma^a_{22} (x_1) = 2 \mu^a \, S(h_a k) \, k \alpha \cos(k x_1), \label{sa}
\eea
and, the stress $\sigma_{22}^b(x_1)$ along the surface of the film
$b$ is 
\bea
\sigma^b_{22} (x_1) = -2 \mu^b \, S(h_b k) \, k \beta \cos(k x_1), \label{sb}
\eea
where $S$ is a nondimensional function defined as
\bea
S(\xi) = \frac{1 + \cosh(2 \xi) + 2 \xi^2}{\sinh(2 \xi) - 2 \xi}. \label{Sfunc}
\eea
For future discussion we note that $S(\xi) \longrightarrow 1/\xi^3$ as
$\xi \longrightarrow 0$, and $S(\xi) \longrightarrow \coth(\xi)$ when $\xi \longrightarrow \infty$.

Substitution of \prn{sa} and \prn{sb} in \prn{sabif} and \prn{sbbif}
leads to the following homogeneous set of equations for $\alpha$ and
$\beta$:
\bea
\pmatrix{ 2 \mu_a k S(h_a k) - Y & Y \cr
          -Y                     & Y - 2 \mu_b k S(h_b k) } \pmatrix{
\alpha \cr \beta} = \pmatrix{ 0 \cr 0} .
\eea
The condition for the existence of nontrivial bifurcation fields,
i.~e., the condition for the existence of nontrivial solutions for
$\alpha$ and $\beta$ is obtained by setting the determinant of the
coefficient matrix to zero resulting in a relation between the interaction
stiffness $Y$ and the wavenumber $k$ of the bifurcation field
\bea
Y = \frac{2 k \, \mu_a \, \mu_b \, S(h_a k) \, S(h_b k)}{\mu_a \,
S(h_a k) + \mu_b \, S(h_b k)} . \label{bifcond}
\eea
If, for a given value of $Y$, there is at least one real value of $k$
that solves \prn{bifcond}, then the homogeneous solution is unstable
and the films deform inhomogeneously. The {\em lowest value} of $Y$ for
which bifurcations are possible is called the {\em critical
interaction stiffness} denoted by  $Y_c$; the wavenumber(s) of the mode(s) that satisfies(satisfy)
\prn{bifcond} for $Y = Y_c$ is(are) called the critical mode(s) and
the(these)
wavenumber(s) is(are) denoted by $k_c$. 

At this point, there are four parameters $\mu_a, \mu_b, h_a, h_b$ that
enter into the determination of the stability of the system. The
discussion to follow is much simplified on introduction of effective
parameters; to this end, we define 
\bea
\mu = \frac{\mu_a \mu_b}{\mu_a + \mu_b}
\eea
and
\bea
h = h_a + h_b.
\eea
We also introduce nondimensional parameters $M$ and $H$ defined as
\bea
\frac{\mu}{\mu_a} = M, \;\;\;\;\;\; \frac{\mu}{\mu_b} = (1-M)
\eea
and
\bea
h_a = H h, \;\;\;\; h_b = (1-H) h.
\eea
When $\mu_a \ll \mu_b$, $M \rightarrow 1$ and $M \rightarrow 0$ if
$\mu_b \ll \mu_a$; similarly when $h_a \gg h_b$, $H \rightarrow 1$,
while $H \rightarrow 0$ implies $h_b \gg h_a$.  Another important
quantity of interest is the effective elastic stiffness $\Ke$ of the
two-film system defined as
\bea
\Ke = \frac{\displaystyle{\frac{\mu_a}{h_a} \,
\frac{\mu_b}{h_b}}}{\displaystyle{\frac{\mu_a}{h_a} +
\frac{\mu_b}{h_b}}} = \frac{\mu_a \, \mu_b}{\mu_a h_b + \mu_b h_a} =
\frac{\mu}{h} \frac{1}{((1-H)(1-M) + HM)}.
\eea

Based on the  above definitions, the expression for the interaction
stiffness in \prn{bifcond} can be recast in nondimensional
form as  
\bea
\frac{Y}{\Ke} = \frac{2\,((1-H)(1-M) +
HM)\,q\,S(Hq)\,S((1-H)q)}{(1-M)\,S(Hq) + M \, S((1-H)q)} , \label{ndbcond}
\eea
where $q = h k$.

The results for the case when $\mu_b \gg \mu_a$ obtained by
\citeasnoun{Shenoy2001} can be immediately recovered. When 
$M \rightarrow 1$, $K_e \rightarrow \displaystyle{\frac{\mu_a}{h_a}}$ and
\bea
\frac{h_a Y}{\mu_a} = 2 \,Hq \, S(Hq) = 2 \,h_a k \, S(h_a k)
\eea 
The critical interaction stiffness can be obtained as $Y_c = 6.22
\mu_a/h_a$ and $h_a k_c = 2.12$, which is precisely the result of
\citeasnoun{Shenoy2001}. Another route to the same result is to let $h_a
\gg h_b$, i.~e., as $H \rightarrow 1$. In this case, $K_e \rightarrow
\displaystyle{\frac{\mu_a}{h_a}}$ again, and
\bea
\frac{h_a Y}{\mu_a} = 2 \, Hq \, S(Hq)  = 2 \, h_a k \, S(h_a k)
\eea
where the property of the function $S$ near 0, noted above, is
used.  It can be easily shown that similar
results are obtained when either $\mu_b \ll \mu_a$ ($M \rightarrow
0$) or $h_b \gg h_a$ ($H \rightarrow 0$). 

In this section, we have derived the condition for the onset of
instability in a system with two interacting incompressible elastic
films. In addition, it is shown that the results of
\citeasnoun{Shenoy2001} who treated a film interacting with a rigid
contactor, are a special case of the present formulation. The more
general case is taken up for discussion in the next section.

\section{Results and Discussion}
\label{Results}

\begin{figure}
\centerline{\epsfxsize=17.5truecm \epsfbox{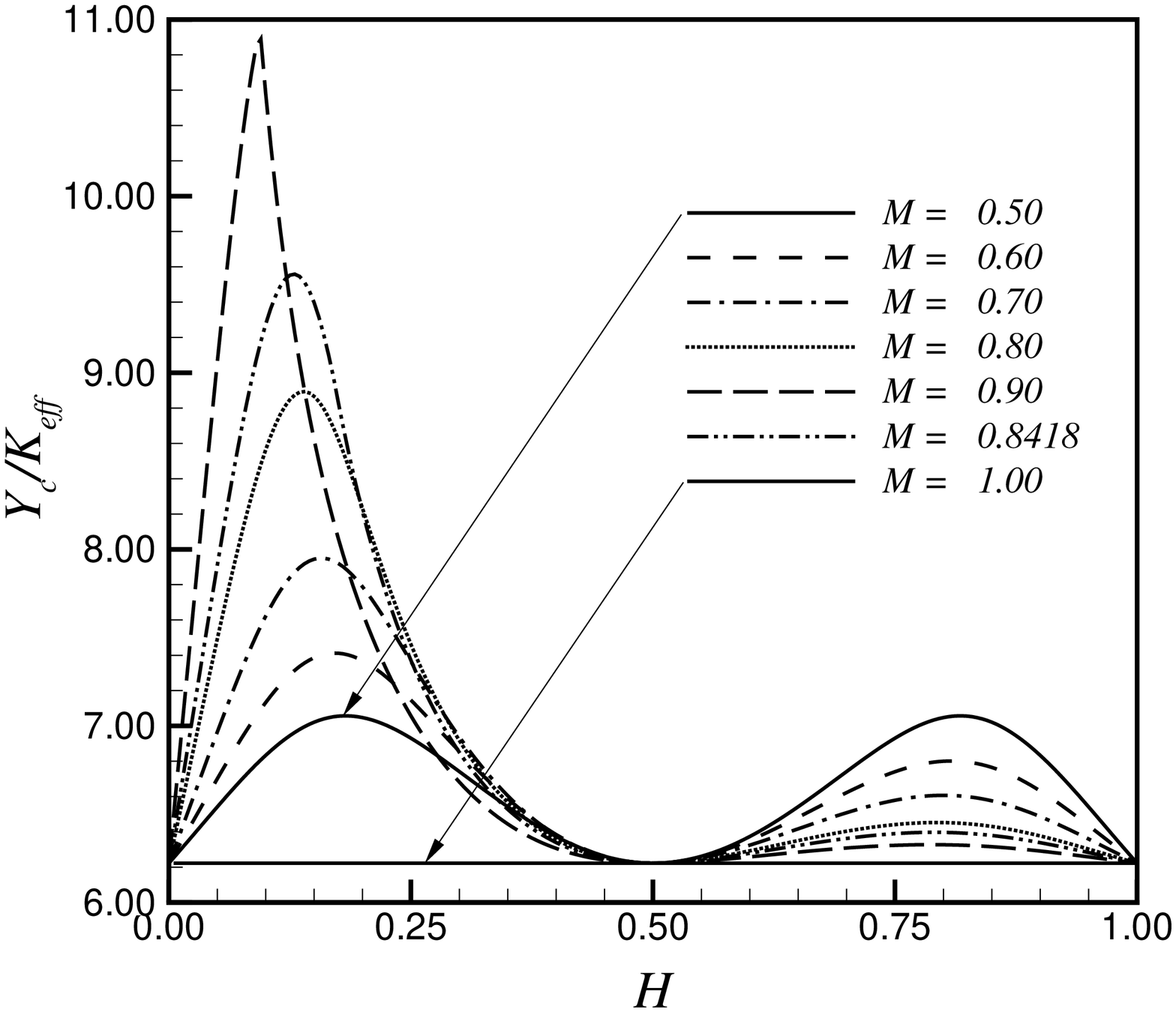}}
\figcap{The critical interaction stiffness as a function of $H$ for
various values of $M$.  }
\label{Yc}
\end{figure}

The nondimensional quantities $M$ and $H$ introduced in the last
section can take any value from 0 to 1. However, it is evident that
the results for a given value of $M$ and $H$ are physically identical
to that of $(1-M)$ and $(1-H)$ (as is evident from 
\prn{ndbcond}). Due to this reason, only the regime $\frac{1}{2} \le M \le
1$ and $0 \le H \le 1$ is considered. Results are discussed in three
categories: (i) $\mu_a = \mu_b$, $h_a \ne h_b$ (ii)  $\mu_a
\ne \mu_b$, $h_a = h_b$ (iii) $\mu_a \ne \mu_b$,  $h_a \ne h_b$.
The results of all these cases are plotted in two sets of graphs shown
in \fig{Yc} and \fig{hakc}. 

An important point about the results is that, just as in the case of a
film interacting with a rigid contactor, the wavelength of the
instability does not depend on the magnitude and nature of the
interaction forces, but depends only on the thicknesses of the films
and their moduli, i.~e., on the parameters $H$ and $M$.

\subsection{Films with Equal Shear Moduli}
\begin{figure}
\centerline{\epsfxsize=17.5truecm \epsfbox{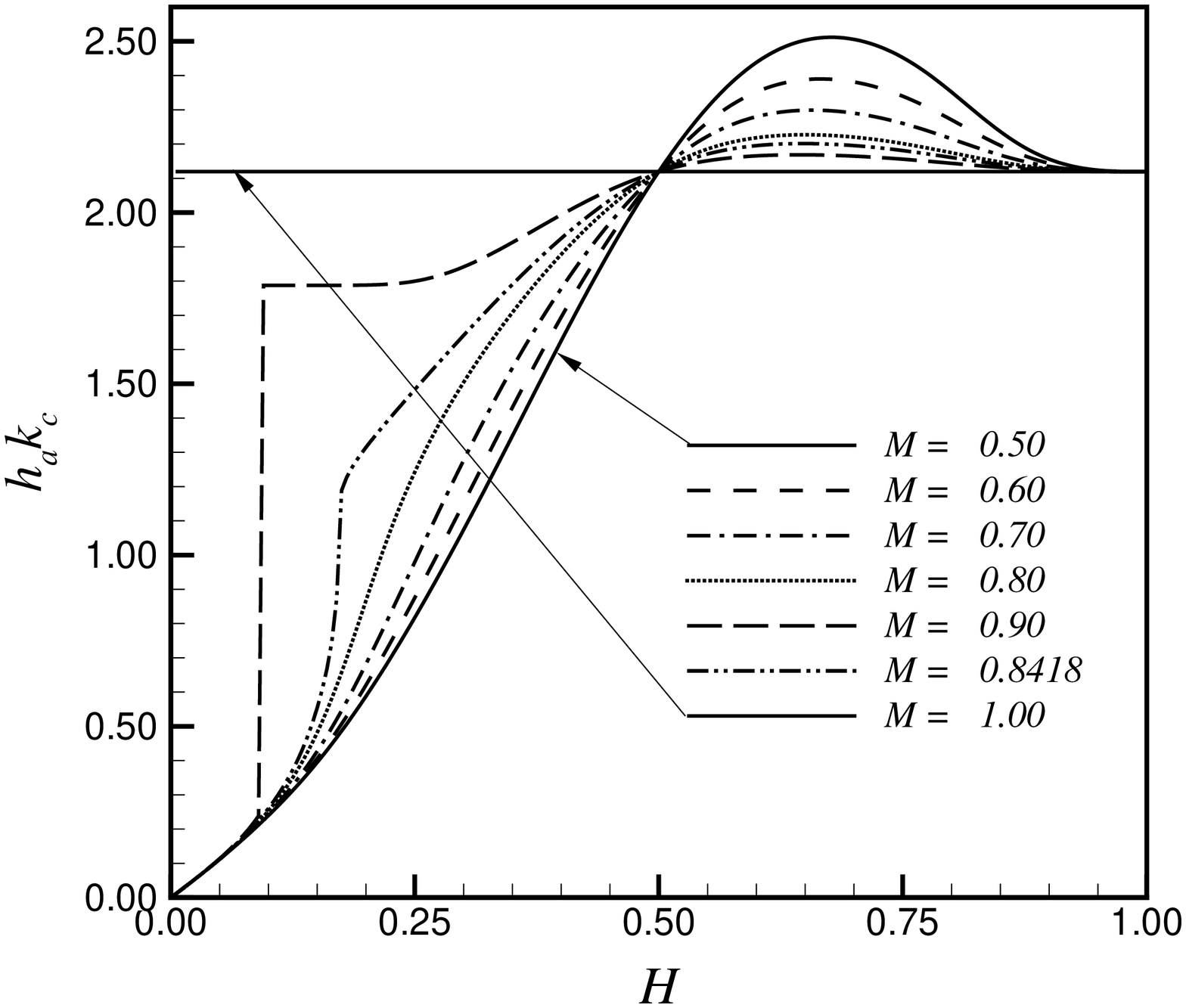}}
\figcap{The wavelength of the critical mode as a function of $H$ for
various values of $M$.}
\label{hakc}
\end{figure}

When the shear moduli of the two films are equal ($\mu_a = \mu_b$), $M
= \half$, $\mu = \mu_a/2$ and $\Ke = H \mu_a/ h_a$. The results for
$Y_c$ and $k_c$ correspond to $M=0.5$ in \fig{Yc} and \fig{hakc}. The
magnitude of  non dimensional critical interaction stiffness 
$Y_c/\Ke$ increases with decreasing $H$ attains a peak and falls as
$H$ approaches $\half$.  If $h_a$ is kept fixed and $h_b$ is changed
to change $H$, then increasing $Y_c/\Ke$ does not imply an increase in
$Y_c$, since $\Ke$ falls linearly with $H$. Thus for a fixed $h_a$,
the effective stiffness of the system reduces with decreasing
$H$. For example, when $H=3/4$, i.e, $h_b/h_a =
1/3$, the value of $K_e = 3\mu_a/4h_a$ and the value of $Y_c/K_e =
6.92$ or $Y_c = 5.19 \mu_a/h_a$ which is {\em less} than $Y_c = 6.22
\mu_a/h_a$ when the film $a$ interacts with a rigid contactor. Looking
at the same example from the point of view of film $b$, we note that
$K_e = (1-H) \mu_b/h_b$ and  $Y_c = 1.73 \mu_b / h_b$ which is
again less than the value $6.22 \mu_b/h_b$. In fact, when films have
equal thicknesses and equal shear moduli, the value of $Y_c = 3.11
\mu_a/h_a$.  The conclusion of these observations is that the
magnitude of the critical interaction stiffness is less than that when
either of the film is interacting with a rigid contactor. Physically,
this can be understood by the fact that introduction of a second film
on the contactor makes the system {\em more compliant}, i.~e., the
effective elastic stiffness of the system comes down. Thus, the elastic
energy penalty required to cause an inhomogeneous deformation in the
system also comes down. As a consequence, the absolute value of the
interaction stiffness required to trigger the instability is reduced. 
Moreover, the actual value of $Y_c$ is governed by the more compliant film.

Turning now to a discussion of the wavenumber of the critical mode, we
note that as $H$ decreases from 1, $h_a k_c$ increases and then falls
subsequently.  For values of $H$ close to 1, the instability triggers
a shorter wavelength mode in film $a$ as compared to when the film $a$
interacts with a rigid contactor.  For example, when $H = 3/4$, the
value of $h_a k_c = 2.45$; in terms the film $b$ the value of $k_c$
is $h_b k_c = 0.82$, i.~e., the wavelength of the
critical mode is much larger than that if the  film $b$ had been
interacting with a rigid contactor. The key point is that the
wavelength of the instability in this regime is {\em intermediate} to
critical wavelengths determined by the thicknesses of the
participating films. As the thicknesses are made equal, i.~e., when $H =
\half$, the critical wavelength becomes $h_a k_c = 2.12$ which is exactly
the wavelength had either of the films been interacting with a rigid
contactor. When $M=1/2$, the case of $H < 1/2$ is
physically identical to the case $H>1/2$. These results can be
interpreted physically as follows. The dependence elastic energy in
the film on the wavenumber $k$ is determined solely be the thickness
of the film, more precisely by the function $S$ (defined in
\prn{Sfunc})  which depends on the
thickness of the film, the shear modulus is a mere multiplicative
parameter. The elastic energy (per unit length of the film) is a minimum when
the wavenumber is equal to $2.12/\mbox{(thickness)}$. However, when
there are two films with {\em very different thicknesses}, the
total elastic  energy in the system (sum of elastic energies in the
two films) can be minimised by choosing an intermediate wavelength.

\subsection{Films with Equal Thicknesses}
When the films are made of equal thickness $h_a = h_b$, the parameter
$H = \half$. For this case, the effective stiffness $K_e = \mu/ h_a$,
and the nondimensional critical interaction stiffness is independent
of the value of $M$, with $Y_c = 6.22 \mu/h_a$. Since $\mu < \mu_a$,
and $\mu < \mu_b$, the actual magnitude of the critical interaction
stiffness is less than the critical interaction stiffness of the films
if they had been interacting with  rigid contactors. This can, again,
be understood by the argument that the effective stiffness of the
system comes down on introduction of the second film on the
contactor. The critical wavenumber, however, does not depend on $M$ as
is evident from \fig{hakc} and is {\em equal} to the wavelength had
the films been interacting with a rigid contactor. Again, this
can understood from the argument stated above that the elastic energy
of a mode is determined by the thickness of the film via the function
$S$ and the moduli are multiplicative constants in the energy
expression. This idea is evident from \prn{ndbcond} on substitution of
$H = \half$.

\subsection{Films with Unequal Thicknesses and Moduli}

The physics of the instability in the general case can be understood
based on the results of the previous two cases. When $M > \half$, the
film $a$ is softer than the film $b$, i.~e., $\mu_a < \mu_b$. When $M$
is close to $1/2$, the qualitative features of the instability are
unchanged. The nondimensional interaction stiffness $Y_c/\Ke$ increases
with decreasing $H$ from 1, and falls as $H$ approaches $\half$. On
further decrease of $H$, $Y_c/\Ke$ increases and decreases again to
$6.22$ as $H$ approaches 0. The behaviour of the critical wavelength
is qualitatively identical to that of the case when $M = \half$.

\begin{figure}
\centerline{\epsfxsize=17.5truecm \epsfbox{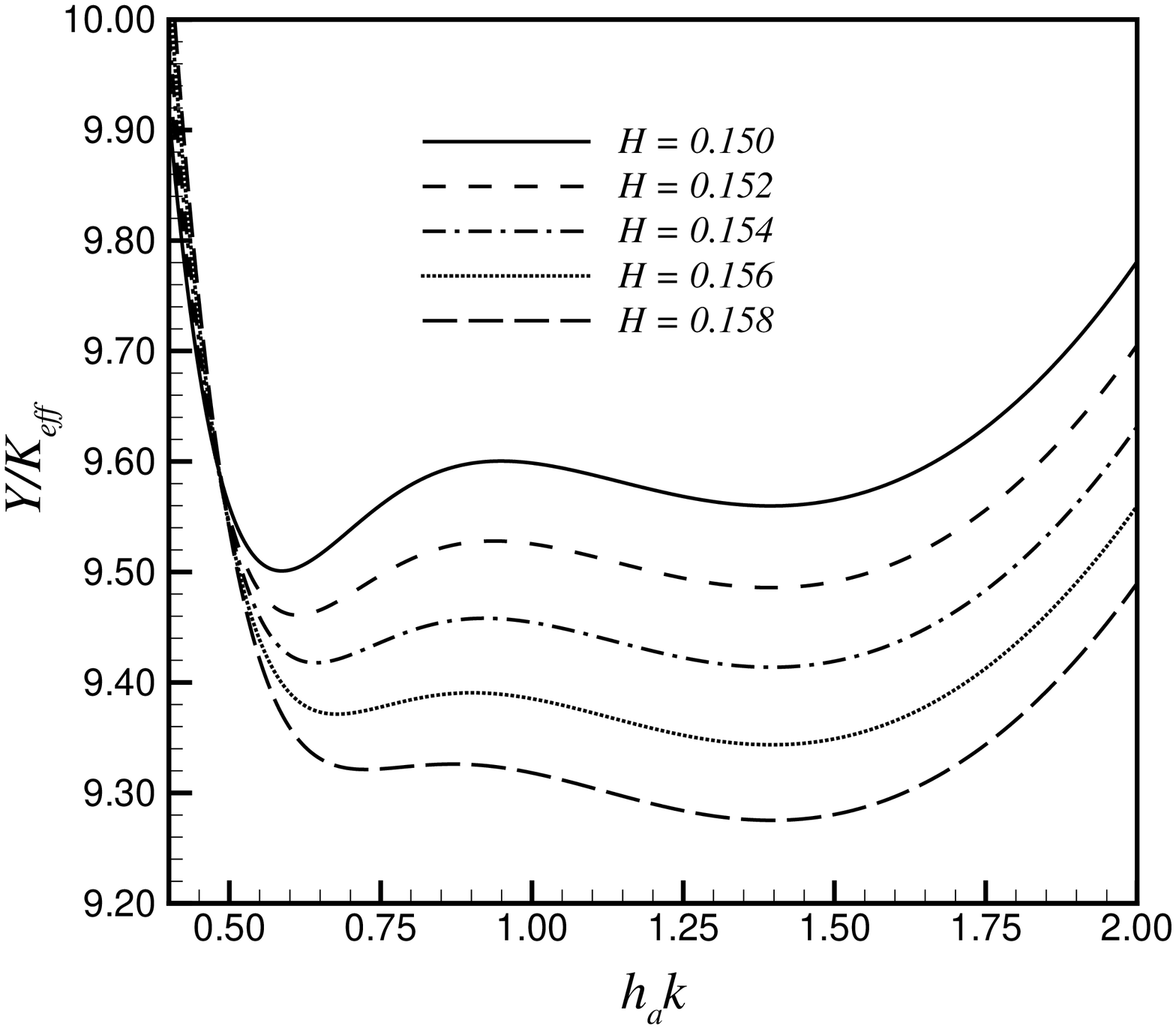}}
\figcap{The interaction stiffness $Y/\Ke$ as a function $h_a k$ showing
multiple minima for different values of $H$. Note how the value of
$h_a k$ at which $Y/\Ke$ attains a minimum changes as $H$ is reduced.}
\label{jump}
\end{figure}

\begin{figure}
\centerline{\epsfxsize=17.5truecm \epsfbox{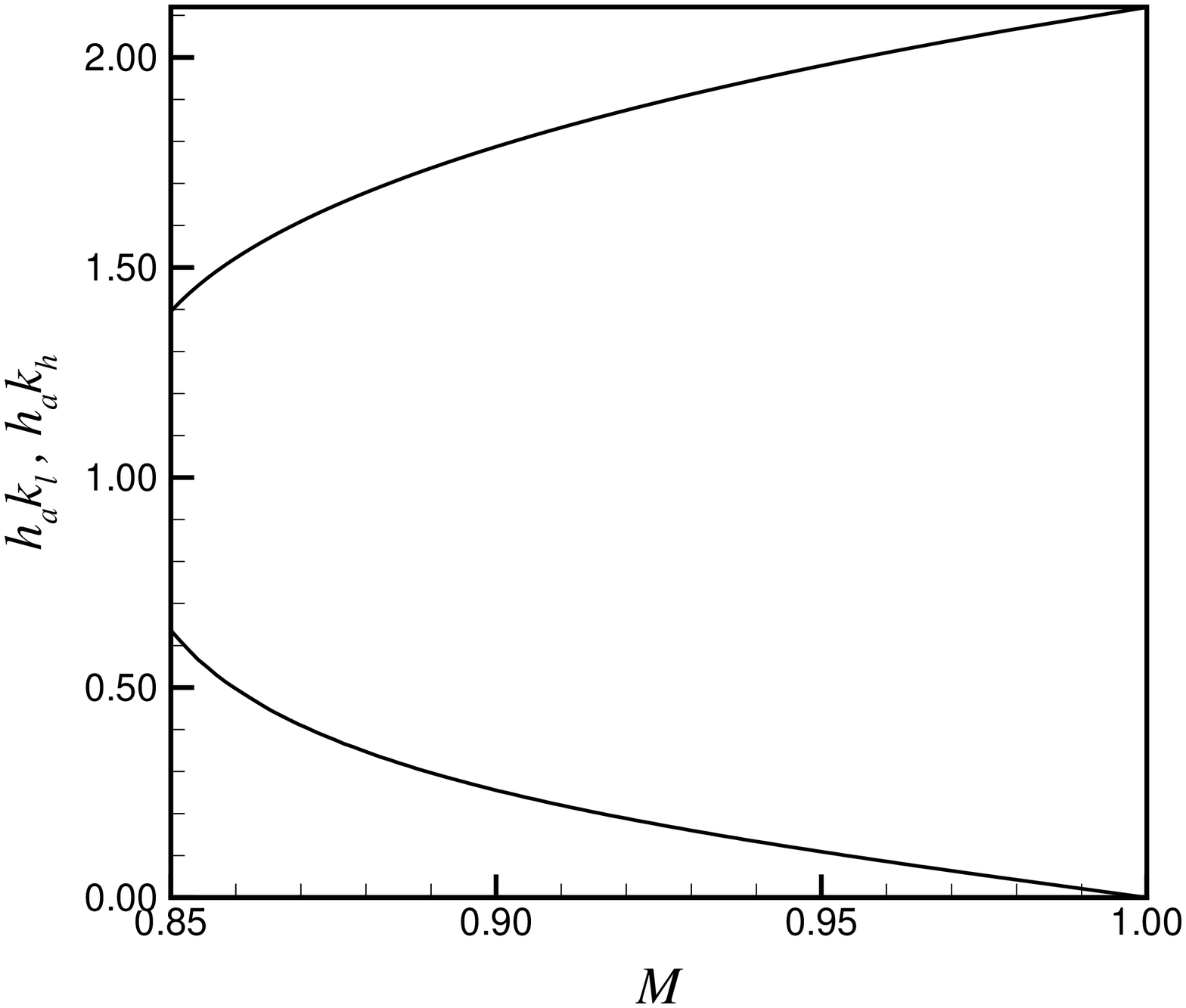}}
\figcap{The lower ($h_a k_l)$ and higher ($h_a k_h$) wavenumbers at $H
= H_J$, as a function of $M$.
}
\label{klkh}
\end{figure}

As $M$ is made close to 1 $(M<1)$, a qualitative change appears in
the solution. In fact, when $M = 0.8418$ (corresponds to $\mu_b/\mu_a =
5.32$), the critical wavenumber changes suddenly as a function of $H$
at about $H = 0.1756$ (corresponds to $h_b/h_a = 4.69$). In fact, for
larger values of $M$, as $H$ is reduced, the critical wavenumber
remains a constant (and close to the wavenumber $h_a k_c =2.12$) but
jumps suddenly to a much lower value -- the value of $H$ at which the
jump occurs is called $H_J$. This is
evident when $M = 0.9$ (shown in \fig{hakc}) where the wavenumber of
the critical mode $h_a k_c \approx 1.79$ when $H < 0.25$; however, at
$H =0.095 = H_J$ (corresponding to $h_b/h_a = 9.52$), the critical wavelength
becomes $ h_a k_c = 0.22 \longrightarrow h_b k_c = 2.1$, i.e., the
wavenumber of the instability changes to the critical wavenumber of
the film $b$ had it been interacting with a rigid contactor. The value
of $H_J$, i.~e., the nondimensional value of $H$ at which the critical
wavenumber jumps from a higher value $k_h$ to a lower value $k_l$
depends on the ratio of the shear moduli; this dependence is plotted
in \fig{Hj}. It is evident that $H_J$ decreases with $M$, as is
expected from the physical arguments presented below. The two
wavenumbers at $H_J$ are plotted in \fig{klkh}. Again, as $M$
approaches 1,  $h_a k_h$
approaches 2.12, the value determined by the thickness of the film
$a$, and $h_a k_l$ approaches 0 (it can, in fact, be shown that $h_b
k_l$ approaches 2.12). 

\begin{figure}
\centerline{\epsfxsize=17.5truecm \epsfbox{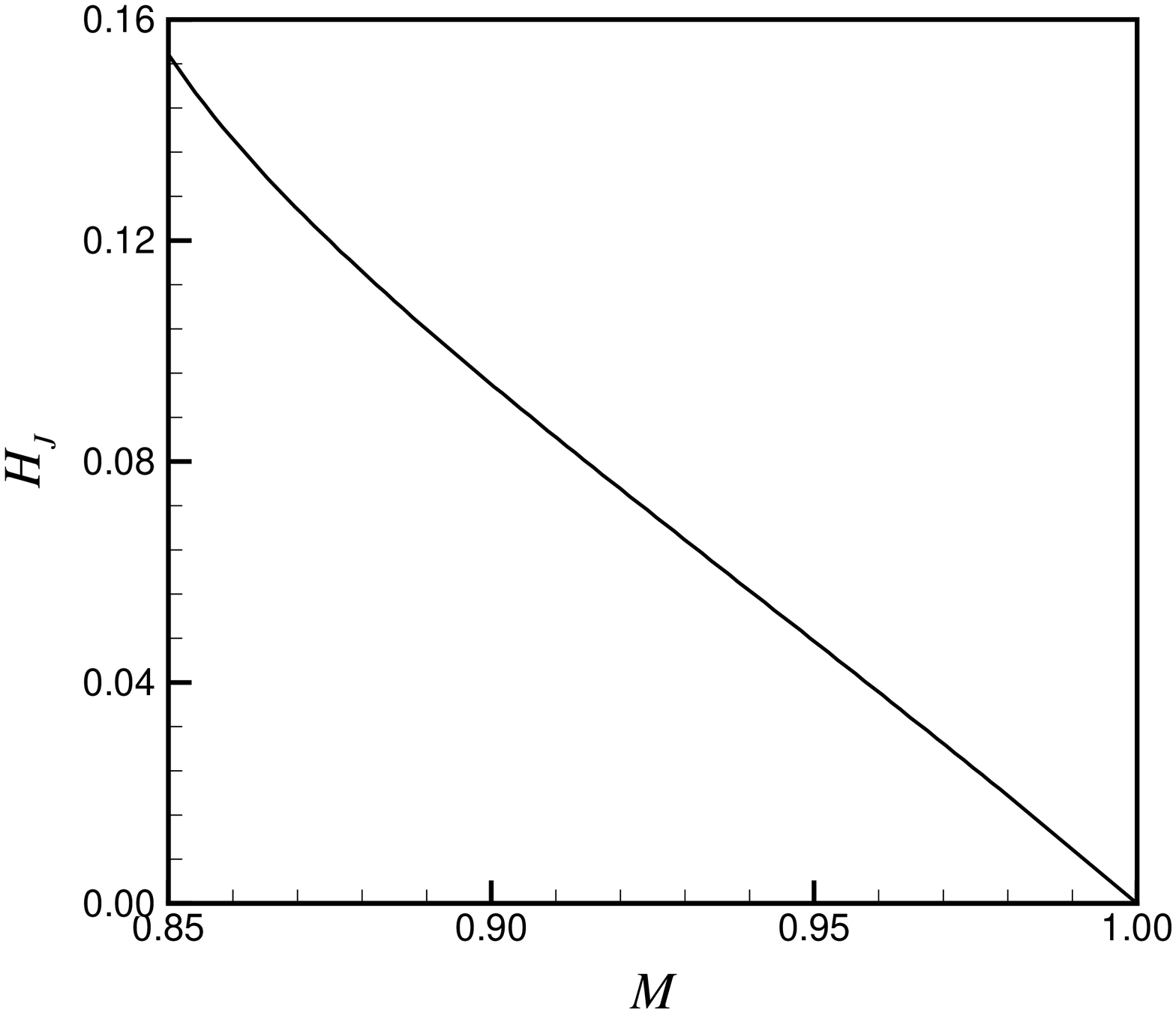}}
\figcap{The nondimensional parameter $H_J$ (that characterises the
difference in the thicknesses of the films) at which the critical
wavenumber jumps, as a function of $M$.
}
\label{Hj}
\end{figure}

From a physical perspective, the total elastic energy is the sum of
elastic energies in the films which are products of their respective
shear moduli with the $S$ function defined in \prn{Sfunc}. When the
elastic moduli are very different, the total energy of the system can
have multiple minima, which implies that $Y$ as a function of $q$
defined in \prn{ndbcond} has multiple minima. Indeed, a plot of $Y$ as
a function of $h_a k$ shown in \fig{jump} clearly shows the presence of
multiple minima; additionally, the value of $h_a k$ where the minimum
is attained changes to a smaller value of $h_a k$ as $H$ is
reduced. The key physical idea is that instability is governed by the
film that has a smaller stiffness (the compliant film). Although
$\mu_b$ is much larger than $\mu_a$,  $h_b$ can be made larger than
$h_a$ to the extent that the stiffness of the film $b$ is smaller than
that of $a$. Thus when $H$ is lowered, the critical wavenumber of the
instability of the system jumps from a wavenumber close to the
critical wavenumber determined by film $a$ (as if interacting with a
rigid contactor) to that of the film $b$  (as if interacting with a
rigid substrate).

\section{Summary and Conclusion}
\label{Summary}
This paper extends the work of \citeasnoun{Shenoy2001}, who studied the
interaction of a soft thin film with a rigid contactor, to the case
where the contactor also has a soft film bonded to it. 
This new analysis uncovers some interesting new results that are summarised below.
\begin{enumerate}
\item{Instability sets in when the interaction stiffness exceeds a
critical value which is determined by the shear moduli and thicknesses
of the interacting films. The magnitude of the critical interaction
stiffness is smaller than that of the critical interaction stiffnesses
of the individual films had they been interacting with  rigid
contactors. Thus it is ``easier'' for the instability to occur in the two-film system.}
\item{In all the cases, the wavelength of the instability does not
depend of the nature and magnitude of the interaction. The critical
wavelength is determined only by the thicknesses and shear moduli of
the films.}
\item{When the films have equal shear moduli, the critical wavenumber of
the instability depends on the relative thicknesses of the
films. When the thicknesses of the films are very different, the
wavelength of the instability is intermediate between the wavelength of
the instability in each of the films had they been interacting with 
rigid contactors. However, when the thicknesses of the film are equal,
the wavelength of the instability is exactly same as that when the
films interact with rigid contactors.  }
\item{When the films have equal thicknesses, but different shear
moduli, the magnitude of the critical interaction stiffness is less
than that of each of the film had they been interacting with rigid
contactors. The wavelength of the critical mode, however, is equal to
that when the films interact with  rigid contactors and is independent of the shear
moduli of the films.}
\item{When the films are of both unequal thicknesses and moduli, the
behaviour is dominated by the less stiff films (stiffness being the
ratio of the shear modulus to the thickness of the film). As long as
the modulus of the contacting film is less than about five times that
of the substrate film, the behaviour of the solution is qualitatively same as
that discussed in the above cases. However, when the modulus of the
contacting film is larger than 5.32 times the thickness of the film on
the substrate, the wavelength of the instability depends strongly on
the thicknesses of the films. As the thickness of the contacting film
is increased, the wavelength of the instability jumps from a value
determined by the thickness of the substrate film to that determined
by the contacting film.  }
\end{enumerate}

It is hoped that the paper will stimulate further experiments along
the lines of \citeasnoun{Ghatak2000} and \citeasnoun{Monch2001} to
verify the results  presented here. The analysis
presented here neglects the effect of surface energies and the
compressibility of the films. The inclusion of these effects along
with the viscosity of the films  will be presented elsewhere 
\cite{Sarkar2001}.

\subsection*{Acknowledgement}
VS wishes to thank DST, India for support of this work under the Fast
Track Scheme.

\appendix
\section{Stresses Along the Surface of an Incompressible Film Bonded
to a Substrate with Sinusoidal Surface Deformation}

The aim of this appendix is to outline the determination of the normal
traction along the surface of an incompressible film bonded to a rigid
substrate, the surface of which is deformed sinusoidally. The main
result of this appendix is the formula used in \prn{sa} and \prn{sb}.

Consider an incompressible elastic film described by coordinates
$(x_1,x_2)$ of thickness $h$ bonded to a rigid substrate such that the
free surface of the film has coordinate $x_2 = 0$ and the interface
between the film and the substrate has coordinate $x_2 = -h$. The film
has a shear modulus $\mu$. The boundary value problem has the
following boundary conditions:
\bea
u_1(x_1,-h) = u_2(x_1,-h) = 0, \label{rigid} \\
\non\\
u_2(x_1,0) = \alpha \, \cos(k \, x_1), \;\;\;\;\;\; \sigma_{12}(x_1,0)
= 0. \label{free}
\eea
The equilibrium equation in terms of the displacements is
\bea
\mu \, u_{i,jj} + p_{,i} = 0, \label{navier}
\eea 
where $p$ is the pressure field. The incompressibility condition is
expressed as
\bea
u_{i,i} = 0. \label{incomp}
\eea
A general solution of the set of differential equations that anticipates
the boundary conditions \prn{rigid} and \prn{free} is
\bea
u_1(x_1,x_2) & = & -\frac{\alpha}{k} \left( (B + k (A + B x_2)) e^{k x_2}
+ (D - k (C + D x_2)) e^{-k x_2} \right) \sin(k x_1) \label{uxgen} \\ 
\non \\
u_2(x_1,x_2) & = & \alpha  \left( (A + B x_2)  e^{k x_2} + (C + D x_2)
e^{-k x_2} \right)\cos(k x_1) \label{uygen} \\ 
\non \\
p(x_1,x_2) &  = & -2 \mu \alpha \left( B e^{k x_2} + D e^{-k x_2} \right)
\cos(k x_1) \label{pgen}
\eea
The constants $A$, $B$, $C$ and $D$ can be determined from the
boundary conditions \prn{rigid} and \prn{free}. The solution is
\bea
A = \frac{ 1 + e^{2 k h} - 2 k h (1 - k h) }{ 2 \sinh(2 kh) - 4 kh},
\;\;\;\;\;\;\; C = 1 - A, \label{Asol} \\
\non \\
B = -\frac{ k ( 1 + e^{2 k h} - 2 k h)}{2 \sinh(2 kh) - 4 kh},
\;\;\;\;\;\;\; D = k + B . \label{Bsol} 
\eea

From these relations, the expression for the normal component of
traction along the surface of the film can be derived as
\bea
\sigma_{22}(x_1,0) = 2 \mu u_{2,2} + p = 2 \mu (2 A - 1) k a \cos(k x_1).
\eea
On substitution of the solution for $A$ from \prn{Asol}, the
expression simplifies to
\bea
\sigma_{22}(x_1,0) = 2 \mu \,  S(h k) \, k\alpha \cos(k x_1)
\eea
where the function $S$ is defined in \prn{Sfunc}.

\bibliographystyle{jmps}
\bibliography{/home/shenoy/Bib/mybib}

\end{document}